\documentclass[
aps,
pra,
amsmath,amssymb,
reprint,
floatfix,
superscriptaddress
]{revtex4-2}

\usepackage{hyperref}
\usepackage{orcidlink}

\usepackage{graphicx}
\graphicspath{{figures/}}
\usepackage{dcolumn}
\usepackage{bm}

\usepackage[utf8]{inputenc}
\usepackage[T1]{fontenc}
\usepackage{braket}
\usepackage{siunitx}
\usepackage{xcolor}

\def\Rb87{^{87}\mathrm{Rb}}                     
\def\K40{^{40}\mathrm{K}}                       

\def\ez{\mathbf{e}_z}

\def\sinc{{\rm sinc}}
\def\fe{f_e}
\def\femax{f_{\rm max}}
\def\deterr{\Delta_{\rm sys}}
\def\DeltaMax{\Delta_d}
\def\output{C} 
\def\Rinv{R^{-1}}
\def\R{R} 
\def\epsilonerr{\delta\epsilon} 
\def\Nerr{\delta N} 

\begin{document}

\title{In situ magnetic-field stabilization for quantum-gas experiments}

\author{E.~Gvozdiovas\ \orcidlink{0000-0003-0782-6705}}
\thanks{These authors contributed equally to this work.}
\affiliation{Institute of Theoretical Physics and Astronomy, Vilnius University, Saulėtekio 3, LT-10257 Vilnius, Lithuania}
\affiliation{Joint Quantum Institute, University of Maryland, College Park, Maryland 20742, USA}
\affiliation{National Institute of Standards and Technology, Gaithersburg, Maryland 20899, USA}

\author{A.~Vald\'es-Curiel}
\thanks{These authors contributed equally to this work.}
\affiliation{Joint Quantum Institute, University of Maryland, College Park, Maryland 20742, USA}

\author{Q.-Y.~Liang\ \orcidlink{0000-0002-2430-7248}}
\affiliation{Joint Quantum Institute, University of Maryland, College Park, Maryland 20742, USA}
\affiliation{Department of Physics and Astronomy, Purdue University, West Lafayette, IN 47907, USA
}
\affiliation{Purdue Quantum Science and Engineering Institute, Purdue University, West Lafayette, IN 47907, USA}

\author{E.~D.~Mercado-Gutierrez\ 
\orcidlink{0000-0001-7143-2618}}
\affiliation{Joint Quantum Institute, University of Maryland, College Park, Maryland 20742, USA}
\affiliation{National Institute of Standards and Technology, Gaithersburg, Maryland 20899, USA}

\author{A.~M.~Pi\~{n}eiro\ \orcidlink{0000-0001-5988-5227}}
\affiliation{Joint Quantum Institute, University of Maryland, College Park, Maryland 20742, USA}
\affiliation{National Institute of Standards and Technology, Gaithersburg, Maryland 20899, USA}

\author{J.~Tao}
\affiliation{Joint Quantum Institute, University of Maryland, College Park, Maryland 20742, USA}
\affiliation{Department of Physics, Harvard University, Cambridge, Massachusetts 02138, USA}

\author{D.~Trypogeorgos \orcidlink{0000-0002-8490-8918}}
\affiliation{CNR Nanotec, Institute of Nanotechnology, via Monteroni, 73100, Lecce, Italy}

\author{M.~Zhao\ \orcidlink{0000-0001-5708-6683}}
\affiliation{Joint Quantum Institute, University of Maryland, College Park, Maryland 20742, USA}

\author{I.~B.~Spielman\ \orcidlink{0000-0003-1421-8652}}
\email[Author to whom correspondence should be addressed: ]{ian.spielman@nist.gov}
\homepage{http://ultracold.jqi.umd.edu}
\affiliation{Joint Quantum Institute, University of Maryland, College Park, Maryland 20742, USA}
\affiliation{National Institute of Standards and Technology, Gaithersburg, Maryland 20899, USA}

\date{\today}

\begin{abstract}
We demonstrate a minimally-destructive in situ technique for measuring and stabilizing slowly-drifting magnetic fields in ultracold-atom experiments.
While conventional magnetic-field sensors such as Hall, giant magnetoresistive, or fluxgate-based devices are broadly used, their accuracy, precision and dynamic range can be limited.
In addition, these sensors are typically positioned at least several centimeters away from the in-vacuum atomic system, as their operation creates perturbing magnetic fields, and their placement is limited by geometric constraints imposed by the vacuum system.
We overcome these issues by using the atomic system itself as a built-in magnetometer.
To that end, we employ a pair of weak measurements to determine the Zeeman splitting--and thereby the magnetic field---of a magnetically sensitive atomic transition.
We provide closed-form expressions quantifying the trade-offs between measurement noise,  dynamic range, and atom loss.
This procedure is demonstrated with ultracold $\Rb87$, weakly measured using partial-transfer absorption imaging.
We then incorporate a Kalman filter to stabilize the magnetic field; this eliminated long-term drift in the ambient field (as high as $\sim$ 70 nT/hr) in exchange for a modest increase in shot-to-shot variability from $\SI{1.8 \pm 0.2}{\nano\tesla}$ to $\SI{2.0 \pm 0.2}{\nano\tesla}$.
\end{abstract}

\maketitle

%
%

\begin{figure*}
    \centering
    \includegraphics[scale=1]{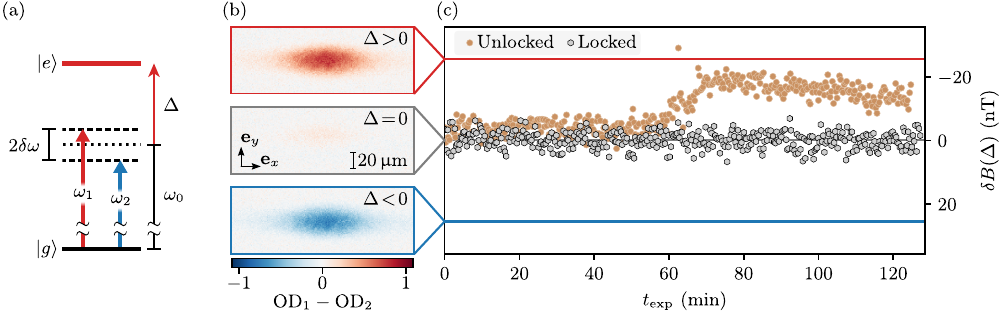}
    \caption{
    (a) Two pulse scheme for the measurement of detuning from resonance $\Delta$.
    Two fixed oscillators $\omega_1$ and $\omega_2$ centered around $\omega_0$ promote atoms from $\ket{g}$ to $\ket{e}$.
    (b) Typical differential in situ partial transfer absorption images from which the feedback signal is obtained at positive, zero, and negative $\Delta$.
    (c) Experimental magnetic field drift $\delta B$ with (gray) and without (brown) the feedback-locking scheme enabled for $B_0 = \SI{80} {\micro\tesla}$.
    }
    \label{fig:intro}
\end{figure*}

\section{Introduction}

The control of magnetic fields is foundational for today's ultracold-atom quantum simulators and computers~\cite{Bloch2008,Wintersperger2023}, for example, facilitating coherent control via precise tuning of Zeeman splittings, optimizing matrix-elements via the natural quantization axis, and controlling interactions via Feshbach resonances~\cite{Chin2010}.
These requirements have motivated substantial hardware development for precise, accurate, and fast magnetic-field control and stabilization~\cite{Dedman2007,Merkel2019,Xiao2020,Rogora2024}.
Despite these advances, determining---and keeping steady---the field at the atoms' position remains challenging in practice.
Environmental noise sources can shift atomic levels and drive unwanted transitions between them.
Such noise leads to dephasing in neutral-atom digital and analog quantum simulators and computers, decreasing their reliability~\cite{Saffman2016,Lin2011}, and
limits the performance of atomic clocks and sensors~\cite{Ludlow2015,Mehlstaubler2018}.
We focus on the particularly vexing case of magnetic-field noise, which couples to internal atomic degrees of freedom via the Zeeman effect, and to external (motional) degrees of freedom through spatial gradients.
In limited cases, magnetic-field sensitivity can be reduced by using magnetic-field-insensitive ``clock transitions'', either native to the atom under study or engineered via dynamical decoupling~\cite{Sarkany2014, Trypogeorgos2018}.
These are not available in most experiments, and so the field itself must be stabilized.

Stable magnetic fields can be created using a variety of approaches.
For example, passive shielding (nested layers of mu-metal) can reduce typical laboratory ambient magnetic fields from the $\SI{}{\micro\tesla}$ scale to the $\SI{}{\nano\tesla}$ regime~\cite{Farolfi2019}.
While passive shielding is ideal for many very low field applications, it can impede optical access and limit the intentional application of large fields.
At higher fields, $\mu$-metals can become magnetized, leading to uncontrolled magnetic field offsets.
In cold-atom settings such as ours, where shielding is impractical, external magnetic fields must be deliberately eliminated.
Cold atom experiments rely on electromagnets driven by low-noise current sources to create dynamic magnetic fields; the required specifications---high bandwidth ($\gtrsim10\ {\rm kHz}$), low noise (part-per-million level), and often high current (up to $1\ {\rm kA}$)---are not available simultaneously from commercial sources, necessitating custom designs
\cite{Yang2019,Borkowski2023,Liu2023,Zhao2023}.
Magnetic-field noise can be further reduced by feed-forward compensation~\cite{ODwyer2020}, which eliminates deterministic ``noise'' sources, and feedback stabilization, which relies on real-time measurements from external sensors~\cite{Duan2022}.

We describe a minimally invasive technique to overcome these limitations by leveraging magnetic-field-sensitive transitions in the ultracold-atom system itself, yielding an in situ magnetometer that provides a feedback signal for magnetic field stabilization. 
Our approach of using the cold-atom system itself as a magnetometer demands simple hardware, and offers very low noise, maximum sensitivity at the desired setpoint, dynamic range matched to external noise sources, and sensing co-located at the atoms.
As detailed below, the primary drawbacks are bandwidth limits, and Dick sampling errors~\cite{Dick1988}.

The vast majority of cold-atom experiments operate in a cyclic manner.
Firstly, cooling down from room temperature (or above) to the ultracold regime is performed on time-scales ranging from just 10's of milliseconds for simple laser-cooling, to many 10's of seconds for creating large quantum gases.
Subsequently, the desired experiment is performed on the atomic ensemble before it is obliterated by measurement, after which the cycle repeats.
Our approach adds a magnetometry stage near the end of this cycle, prior to performing the desired experiment~\cite{LeBlanc_2013,Ana2017,PhysRevLett.116.200402}.

The basic operational principle of this stage is schematically illustrated in Fig.~\ref{fig:intro}(a), beginning with the atomic ensemble prepared in a long-lived internal state $\ket{g}$.
An oscillatory control field of fixed amplitude and duration, with angular frequency $\omega_1$, transfers a small fraction of the ensemble to an excited state $\ket{e}$, after which the transferred atoms are measured.
The same process is then repeated using a second frequency $\omega_2$.
Intuitively, the closer-to-resonance pulse transfers a larger fraction of atoms; as a result, their difference signals the detuning from resonance.
Three such differences are visualized in Fig.~\ref{fig:intro}(b), ranging from red-detuned (top), resonant (middle), to blue-detuned (bottom).
This fractional difference provides an error signal that we feedback on to stabilize the magnetic field.
Panel (c) shows the excursion of the magnetic field from resonance over the course of a few hours with (gray) and without (brown) feedback.
The unlocked data exhibits slow drift (up to $\sim 70$ nT/hr), while the locked data is stable at the individual nT scale.

This technique is particularly powerful because the detuning is probed at the exact location of the atomic ensemble, and temporally just prior to the target experiment.
Still, the in-principle response-time is limited by the combined duration of the transfer-pulses.
This duration, up to a few milliseconds in our experiments, is not the practical limiting factor; the 10's of milliseconds image acquisition time provides a hardware-level response-time-limit.

In our experiments, the uncertainty in the field from a single observation ($\approx 0.8\ {\rm nT}$) is comparable to the short-time magnetic field noise in the lab ($\approx 2\ {\rm nT}$), therefore necessitating an additional low-pass filter to avoid adding noise. 
In this demonstration, we therefore only compensate long-term magnetic field drift on time-scales much slower than the experimental cycle time. 
Because the magnetic field occasionally undergoes unexpected large excursions (for example, from moving nearby magnetized objects), we implemented a reset mechanism to automatically adjust the time constant of the low pass filter.

The remainder of the manuscript proceeds as follows.
Section~\ref{Section_Magnetometry} develops a two-level model for a magnetically sensitive transition, derives closed-form expressions for a two-pulse scheme with trade-offs between sensitivity and dynamic range, then defines the differential error signal and controller gain.
Section~\ref{Section_Experiment} implements the scheme for $\Rb87$ using a microwave hyperfine transition, and benchmarks locking performance with Ramsey interferometry with a discrete-time Kalman filter engaged.
The main text concludes with an outlook and discussion in Section~\ref{sec:conclusion}.
Appendices~\ref{microwave_transitions}--\ref{app:depletion} compile Breit--Rabi sensitivities and quantify the systematics of ensemble depletion.

\section{Partial transfer magnetometry}
\label{Section_Magnetometry}

We obtain the magnetic field by probing the Zeeman shift of a magnetically sensitive state from a reference resonance condition.
This section describes how to use the basic properties of two-level systems to obtain an optimized detuning-signal. 

Here, we focus on an effective two-level system, in practice consisting of a pair of hyperfine ground states of the $\Rb87$ atom (with $\ket{g}$ and $\ket{e}$ in the $F=1$ and $F=2$ manifold, respectively), coupled by a microwave source at a known frequency.
The population in $\ket{e}$ is selectively imaged using a probe laser tuned to the $F=2$ to $F'=3$ transition~\cite{ramanathanPartialtransferAbsorptionImaging2012,serokaRepeatedMeasurementsMinimally2019a,mordiniSingleshotReconstructionDensity2020a}, a technique known as partial transfer absorption imaging ({PTAI}).
See Sec.~\ref{Section_Experiment} for further experimental details.

\subsection{Two level system}

We begin by reviewing a two-level atom subject to an oscillatory coupling field with angular frequency $\omega$ and strength $\Omega$; this configuration is described by the rotating wave approximation Hamiltonian 
\begin{equation} \label{eq:H_TLS}
\hat{H} = \hbar \Delta \ket{e} \bra{e} + \hbar \Omega/2 \ket{g} \bra{e}  + \text{H.c.}
\end{equation}
with detuning $\Delta \equiv \omega_{eg}-\omega$ dependent on the energy gap $\hbar \omega_{eg}$ between the ground $\ket{g}$ and excited $\ket{e}$ states.
Since our focus is on magnetic field sensitive transitions, $\omega_{eg} \equiv \omega_{eg}(B)$ and therefore $\Delta(B)$ are understood as functions of the magnitude $B$ of the applied magnetic field ${\bf B}$.

We consider a system of $N$ atoms initially prepared in $\ket{g}$, which experience the coupling field for a pulse of duration $t$.
The time-dependent Schrödinger equation predicts the fraction of atoms
\begin{align} 	
	\fe \equiv \frac{N_e}{N} &= \frac{\Omega^2}{\Omega^2\!+\!\Delta^2} \sin^2 \left(\frac{t}{2} \sqrt{\Omega^2\!+\!\Delta^2} \right)\label{eq:Rabi_oscillations}
\end{align}
transferred to $\ket{e}$ to be an oscillatory function of time, with generalized Rabi frequency $\sqrt{\Omega^2+\Delta^2}$.
To simplify our analytic treatment, we focus on the case of small pulse area ($t \Omega \ll 1$), for which Eq.~\eqref{eq:Rabi_oscillations} reduces to the standard Fourier-limited sinc~\footnote{We use the standard (non-normalized) sinc function $\sinc{(x)} = \sin{(x)}/x$, suitable for expressions involving angular frequencies.} expression
\begin{align}
	\fe &\approx \left[\frac{t \Omega}{2} \sinc\left(\frac{t \Delta}{2}\right)\right]^2\label{eq:sinc_oscillations}
\end{align}
with maximum transferred fraction $\femax = (t \Omega / 2)^2$ (see Appendix~\ref{app:depletion} results valid for finite $\femax$).

Since our aim is to extract the detuning, we now find the conditions where Eq.~\eqref{eq:sinc_oscillations} is most sensitive to $\Delta$ by maximizing $|\partial \fe / \partial \Delta|$.
This occurs when $\partial^2 \fe / \partial \Delta^2=0$, giving a transcendental equation
\begin{equation}
6 + [(t \DeltaMax)^2 - 6] \cos{ \left(t \DeltaMax \right)} - 4 t \DeltaMax \sin{ \left(t \DeltaMax \right)} = 0 \, ,
\end{equation}
with global maxima in sensitivity at
\begin{equation} \label{eq:Delta_p_t_p}
|t \DeltaMax| \approx 2.61 .
\end{equation}
Given that this is very close to the half-width at half-max (at $t \Delta_{\rm HWHM} \approx 2.78$), it is unsurprising that the corresponding transferred fraction is $\approx0.55\times \femax$.

The small-transfer limit introduced above is also the most experimentally relevant, since the disturbance of the system is an increasing function of $\fe$ independent of the measurement strategy.
We therefore express our control parameters in terms of the peak transferred fraction $\femax$, along with the detuning $\Delta$ and pulse time $t$.
Because Eq.~\eqref{eq:Rabi_oscillations} depends on $\Delta$, a single observation of $\fe$ (with all other parameters fixed) provides some information regarding detuning.
The sensitivity $|\partial \fe / \partial \Delta|$ has an overall scaling $t \femax$, whereby increasing the pulse duration increases the sensitivity to changes in $\Delta$.
At the same time, the overall width of $\fe(\Delta)$ is proportional to $t^{-1}$, so increasing $t$ correspondingly decreases the dynamic range.

There are two added limitations associated with determining $\Delta$ from $\fe$.
First, the sign of $\Delta$ cannot be determined because $\fe$ is an even function with respect to detuning.
Second, because Eq.~\eqref{eq:Rabi_oscillations} is an oscillatory function of $\Delta$,  there is a wide range of parameters for which $\fe(\Delta)$ does not have a unique inverse.

\subsection{Two pulse scheme}\label{sec:two_pulse}

\begin{figure}
\begin{center}
\includegraphics[scale=1]{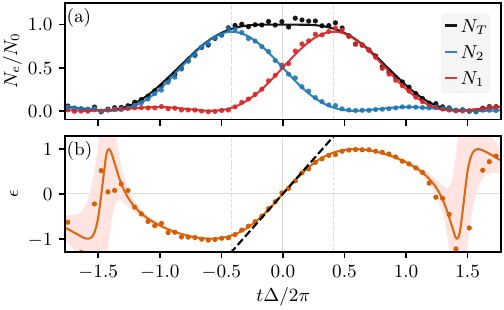}
\caption[]{
Two-pulse PTAI-magnetometry.
(a) Atom number $N_{1,2}$ transferred by each pulse (red, blue) and their sum $N_T$ (black),
(b) the error signal $\epsilon$ vs detuning $\Delta$.
Here $\delta\omega=\DeltaMax$ (dashed lines) and $\femax \approx 0.02$.
Each data point was averaged from 3 experimental measurements with $t=\SI{90}{\micro\second}$.
The slope (black dashed line) is the dimensionless responsivity $R/t$.
Solid lines are obtained from the two level model including finite $\femax$.
The shaded area indicates the uncertainty $\epsilonerr$ given by Eq.~\eqref{eq:sigma_epsilon} for $\delta N/N_0 \approx 0.01$.
}
\label{fig:twopulse}
\end{center}
\end{figure}

Both of these issues can be resolved by measuring $N_e$ twice with two different oscillator frequencies [Fig.~\ref{fig:twopulse}(a)]: more of the ensemble is transferred by the pulse that is closer to resonance, and thus $\Delta$ can be obtained from the relative difference between the two observations. This two-pulse scheme employs the oscillator frequencies
\begin{equation} \label{omega_def}
\omega_1 = \omega_0 + \delta\omega \quad \text{and} \quad \omega_2 = \omega_0 - \delta\omega \, ,
\end{equation}
giving detunings $\Delta_j \equiv \omega_{eg} - \omega_j$ for $j=1,2$; each pulse correspondingly excites an atom number $N_j$ as given by Eq.~\eqref{eq:Rabi_oscillations}.
We define the two-pulse error signal
\begin{equation} \label{epsilon_definition}
\epsilon = \frac{N_1-N_2}{N_T}
\end{equation}
as their difference, normalized to the total transferred number $N_T = N_1 + N_2$.
The error signal $\epsilon$ is independent of the initial ensemble size $N$ and primarily driven by $\Delta \equiv \Delta_0 = \omega_{eg} - \omega_0 $; thus, the latter can be extracted by inverting Eq.~\eqref{epsilon_definition} near $\Delta \approx 0$, where the response to detuning is monotonic and single-valued.

The basic features of this two-pulse scheme are illustrated in Fig.~\ref{fig:twopulse}, with $\delta\omega = \DeltaMax$ set for maximum sensitivity.
In each panel, the predictions of the two level model (solid curves) are in good agreement with experimental data (markers).
First, panel (a) plots the number of atoms $N_e$ excited by each microwave pulse $N_1$ (red), $N_2$ (blue), along with the total number transferred $N_T$ (black).
Here, each $N_e$ is plotted as a fraction of the number transferred at zero detuning $N_0 = N_T(\Delta=0)$; this is a good proxy for the overall number of atoms sacrificed when the detuning is stabilized, with $\Delta \approx 0$. 
We also introduce $f_0 = N_0 / N$ as the fractional transfer at zero detuning.

Next, Fig.~\ref{fig:twopulse}(b) shows the error signal is approximately linear until $|\Delta| \approx \delta\omega$ (dashed lines), and the oscillatory nature of the sinc functions contribute zero-crossings at larger $\Delta$.

The colored band is obtained by error-propagating Eq.~\eqref{epsilon_definition}, which yields
\begin{equation} \label{eq:sigma_epsilon}
(\epsilonerr)^2 = 4\frac{(N_1 \Nerr_2)^2 + (N_2 \Nerr_1 )^2}{N_T^4} \approx 2 \left(\frac{\Nerr}{N_T}\right)^2 \, ,
\end{equation}
computed assuming that detector noise (i.e. photoelectron shot noise) is the leading contributor in the uncertainties $\Nerr_1$ and $\Nerr_2$ corresponding to each measurement.
The right-hand side approximates $N_1=N_2=N_T/2$ and $\Nerr=\Nerr_1=\Nerr_2$.
The uncertainty $\epsilonerr$ increases at large $\Delta$, owing to the overall reduction of detected atoms $N_T$.

Equation~\eqref{eq:sigma_epsilon} only describes the statistical uncertainty; systematic errors are also possible.
For example, a persistent offset in the measured atom number $N_{1,2}$ can significantly alter the error signal $\epsilon$.
In the following, we assume systematic errors are negligible.

\subsection{Parameter selection}\label{sec:parameters}

\begin{figure}
\begin{center}
\includegraphics[scale=1]{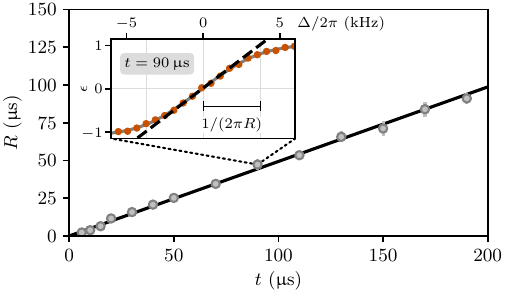}
\caption[]{Responsivity $R$ as a function of pulse duration $t$ with experimental data (markers) plotted along with our model prediction (curve) for $\femax \to 0$.
These data were taken with $\delta\omega = \DeltaMax$, i.e., $t \delta\omega \approx 2.61$.
Inset: $R$ is the linear component (dashed line) of a fifth order polynomial fit (solid curve visible under the data points) to measurements of $\epsilon$ taken as a function of $\Delta$.
The reported uncertainties from the fit are used to generate the error bars in the main panel.
}
\label{fig:R}
\end{center}
\end{figure}

The character of the error signal $\epsilon$ depends strongly on the choice of the oscillator spacing $\delta\omega$.
Here, we pay particular attention to the behavior relevant for magnetic field locking: near $\Delta=0$, with sacrificed atom number $N_0$ held constant.
In this case, the magnetometer can be characterized by its responsivity defined by $\R \equiv \partial \epsilon / \partial \Delta$ (larger is better) and the uncertainty $\epsilonerr$ determined by noise in $N_{1,2}$ (smaller is better).

As shown in the inset to Fig.~\ref{fig:R}, we experimentally obtain $R$ by measuring the slope of $\epsilon$ versus $\Delta$ near $\Delta=0$; in practice, we focus on the monotonic regime (markers), and identify $\R$ as the linear part (dashed line) of a fifth order polynomial fit (solid curve).
Figure~\ref{fig:R} plots the responsivity obtained in this way (markers) with pulse spacing $\delta\omega = \DeltaMax$ [recall from Eq.~\eqref{eq:Delta_p_t_p} that $\DeltaMax\propto t^{-1}$];
these measurements are in good agreement with the small-transfer model's prediction of
\begin{equation}
\R \approx 3.1 t / (2\pi)
\label{eq_R_small}
\end{equation}
as shown by the solid curve, with responsivity improving in proportion to pulse duration.
In what follows, we remove this overall effect by employing the dimensionless ratio $R / t$; similarly, the impact of pulse spacing is best quantified by the dimensionless product $t \delta\omega$.

\begin{figure}[tb]
    \centering
    \includegraphics[scale=1]{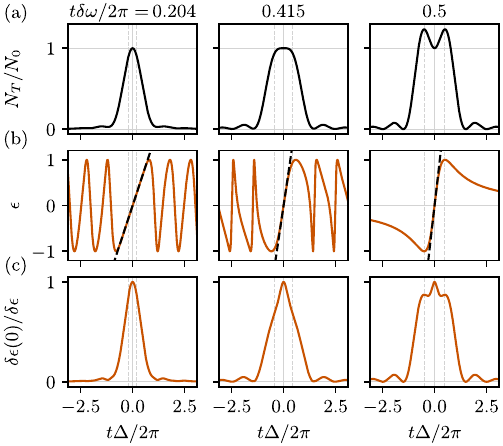}
    \caption{
    Impact of pulse spacing $t \delta \omega$ in the small transfer limit for various $t \delta \omega$.
    (a) Total number transferred $N_T$ as a fraction of the zero detuning transfer $N_0$. 
    (b) Error signal $\epsilon$.
    (c) The inverse of the error signal's uncertainty $\epsilonerr$ as a fraction of its zero-detuning value $\epsilonerr(0)$.
    The vertical dashed lines indicate the pulse frequency spacing $\pm \delta\omega/(2\pi)$.
    }
    \label{fig:regimes}
\end{figure}

Figure~\ref{fig:regimes} shows the overall impact of pulse spacing $t\delta\omega$ ranging from $0$ to $\pi$, starting with the normalized number of sacrificed atoms $N_T / N_0$ as a function of detuning (a). 
For small $\delta\omega$ (left), the $\sinc^2$ peaks from each individual pulse overlap and sum to a single sharp peak; by contrast, for large enough $\delta\omega$ (right), the underlying  $\sinc^2$ functions are well resolved, yielding a pair of peaks in $N_T$.
The optimal pulse spacing $t \DeltaMax$ is intermediate between these and yields a broadened single peak.
Next, Fig.~\ref{fig:regimes}(b) shows example error functions (solid curves) along with the small-$\Delta$ linear approximation (bold dashed lines).
The dimensionless responsivity $\R/t$ given by the slope of these lines increases with $t\delta\omega$.
Lastly, Fig.~\ref{fig:regimes}(c) plots the inverse uncertainty of the error signal $1/\epsilonerr$ normalized to the zero-detuning value $1/\epsilonerr(0)$.
Equation~\eqref{eq:sigma_epsilon} shows that $\epsilonerr(0)$ depends on pulse duration only via the sacrificed number $N_0$, which, as noted above, is held constant.
Similarly to $N_T$, $1/\epsilonerr$ exhibits a peak at zero detuning, that widens with increasing $\delta\omega$.

In addition to these general observations, the specific values of $t \delta\omega$ in Fig.~\ref{fig:regimes} are particularly noteworthy, suggesting different regimes of closed loop operation.

\noindent
{\it Linearity} --- The pulse splitting $t \delta \omega / (2 \pi) \approx 0.204$ (left) was selected so that $\partial^3 \epsilon/\partial \Delta^3=0$ at $\Delta=0$, making the error function exceptionally linear near  $\Delta=0$.
This improves the performance of standard feedback loops---such as PID---which achieve optimal performance for error signals that depend linearly on the control signal (here causing a change in detuning).

\noindent
{\it Continuity} --- When $t \delta \omega / (2 \pi) = 0.5$ (right), the zeros of the underlying sinc functions coincide, making $\epsilon$ a continuous function that always yields an error signal of the correct sign.
In this case, feedback can always direct the system in the direction of resonance, mitigating out-of-lock issues.

\noindent
{\it Number stability} --- By construction, setting $\delta\omega = \DeltaMax$ (center) makes $N_T$ drop only as $\propto \Delta^4$ near $\Delta = 0$.
This choice minimizes the change in remaining atom number after the measurement process.
In order to maximize the number-stability of our final Bose-Einstein condensates (BECs) in closed-loop operation, we select this operating condition.

The preceding discussion assumes the typical operating conditions where $\femax \ll 1$; the weak dependence of all relevant quantities on $\femax$ is explored in Appendix~\ref{app:depletion}.

\section{Closed loop operation}
\label{Section_PID}

Having identified optimal parameters for sensing, we turn to the task of employing this information to stabilize the error signal near zero, i.e., keeping $\Delta \approx 0$ for an extended duration.
As discussed in the introduction, our cold atom experiment operates in a cyclic manner with a $\gtrsim 10\ {\rm s}$ period, where the $j$-th experimental cycle produces a single observation of the error signal $\epsilon_j$.
The feedback protocol described below uses this information to adjust the detuning (i.e., magnetic field) in cycle $j+1$.

The feedback controller employs the discrete-time PI~\footnote{We used a PI loop rather than PID loop both because the delay between measurements is much larger than the response time of our system (i.e., no inertia), and also to avoid noise amplification by differentiation.} expression
\begin{equation} \label{PID_vanilla}
\output_{j+1}\!=\!\ P \epsilon_j + I \sum_{j^\prime=0}^{j} \epsilon_{j^\prime} \!=\!\output_{j} + \left[P\left(\epsilon_j - \epsilon_{j-1}\right) + I \epsilon_j\right],
\end{equation}
where $P$, $I$ are the proportional and integrator constants, which determine the control signal $C_{j+1}$ for use in the next cycle.
As discussed in Sec.~\ref{sec:parameters}, near $\Delta=0$ the lock signal is approximately linear with $\Delta \approx \Rinv \epsilon$.
Therefore, when $P=0$ and $I=-\Rinv$, the second expression in Eq.~\eqref{PID_vanilla} reduces to the ideal correction required to bring the system back to resonance.

\subsection{Kalman filter} 
\label{sec:noise_aware}

We now add to our discussion the unavoidable impact of measurement noise.
In reality, the measured error signal $\epsilon_j$ has uncertainty $\delta \epsilon_j$ from a statistically uncorrelated combination of measurement noise $\delta \epsilon^{(m)}_j$ and detuning noise $\delta \epsilon^{(d)}$.
For example, the ideal correction with $P=0$ and $I=-\Rinv$ can be seen to directly write noise into the feedback signal $C_{j+1}$.
In this limit, reducing $I$ decreases the injected noise in exchange for only partly correcting the detuning error.
In the remainder of this section, we introduce a Kalman-like feedback loop~\cite{Kalman1960} that maximizes feedback bandwidth while minimizing added noise.

The essential idea of any Kalman filter is to maintain a running model of the expected signal $m_j$ and uncertainty $\delta m_j$ based on the past history of measurements and knowledge of the system dynamics.
Here the only deterministic system dynamics result from the feedback operation itself.
Therefore, an update of the model (and analogously the uncertainty) from timestep $j$ to $j+1$ proceeds as follows: (A) the model begins with an estimate of the system state $m_j$; (B) this is revised to $m'_j$ based on the measurement result $\epsilon_j$; finally (C) the model predicts $m_{j+1}$ using Eq.~\eqref{PID_vanilla} and the responsivity $R$.
For example, when $P = 0$, the last step predicts $m_{j+1} = (1 + I_j \R) m'_j$ and selecting $I_j = -\Rinv$ leads to $m_{j+1} = 0$ after every feedback operation.
That is to say, like many scientists, the model believes itself to be perfect.

Given this overview, we now expand on each of these steps.
At timestep $j$ we incorporate the information provided by the new measurement using the average
\begin{align}\label{eq:kalman}
m'_j &= \frac{(\delta \epsilon_j)^{-2} \epsilon_j + (\delta m_j)^{-2} m_j}{(\delta \epsilon_j)^{-2} + (\delta m_j)^{-2}}
\end{align}
weighted by the inverse-variances; the model's expected variance is itself updated via 
\begin{align} \label{eq:kalman_sigma}
\delta m^2_{j+1} &= \frac{N_{\rm eff} \delta \epsilon^2_j+(\epsilon_j- m_j)^2}{ (N_{\rm eff} + 1)^2 }
\end{align}
with an effective number of measurements $N_{\rm eff} = \left[\epsilonerr_j / \delta m_j \right]^2$.
Updating the model's variance in this way provides a reset mechanism when a measurement outcome is very far from the model's expectation.

In principle, the feedback coefficient could be optimized by 
\begin{align}
I_j &= -\Rinv \times {\rm min}\left(\sqrt{f} \frac{\delta \epsilon^{(d)}}{\delta m_j}, 1\right)
\end{align}
so that the noise from feedback only increases the detuning noise-variance by no more than a user defined fraction $f$.
In our demonstration below, we opt for simplicity set $I_j = -\Rinv$.

We initialize this algorithm at $j=1$ by setting $m_1 = \epsilon_1$ and $\delta m_1 = \delta \epsilon_1$.
Altogether, this realizes a moving average that reduces the influence of measurement noise as more data is accumulated, at the same time robust to disturbances thanks to the built-in reset.

In practice, we use a two-step process to lock the magnetic field.
We start with the simple PI approach [Eq.~\eqref{PID_vanilla}] to benefit from its fast response time; once the lock is established ($\epsilon \approx 0$), we switch to the Kalman filter.

\section{Experimental realization}
\label{Section_Experiment}

The following outlines the aspects of our experimental time sequence as it pertains to the project at hand (see Ref.~\onlinecite{Lin.Rapid.PhysRevA.79.063631} for further information).
Following $10.4\ {\rm s}$ of laser cooling and evaporative cooling, our story begins with $\Rb87$ atoms in the hyperfine $\ket{F=1,m_F=-1}$ ground state, confined in a crossed optical dipole trap formed by a pair of $1064\ {\rm nm}$ laser beams.
A bias magnetic field ($B_0 \approx \SI{80}{\micro\tesla}$) in the direction of gravity $\ez$ lifts the ground state degeneracy between $m_F$ sub-levels: the ideal setting for our magnetometry scheme.

Magnetometry operates best in the monotonic regime of the error function $\epsilon$, so $\delta\omega$ must be selected to exceed the largest plausible shot-to-shot change in detuning; for our lab environment, this motivates the selection $\delta\omega = 2\pi\times 4.6\ {\rm kHz}$.
Assuming the default pulse spacing with $\delta\omega=\DeltaMax$, Eq.~\eqref{eq:Delta_p_t_p} determines the pulse time to be $t = \SI{90}{\micro\second}$, and  Eq.~\eqref{eq_R_small} yields the controller gain $|I| = R^{-1} \approx 2\pi\times3.6\ {\rm kHz}$.
The {PTAI}-magnetometry microwave pulses were both calibrated to transfer $\femax \approx 0.02$ of the ensemble to $\ket{2,-2}$.
After each microwave pulse, we image the atoms transferred to $F=2$ (\SI{30}{\micro\second} exposure time) with resonant absorption imaging on a CMOS camera.
The imaged atoms are ejected from the trap due to the radiation pressure from the probe beam; the remaining atoms in $F=1$ are undisturbed.
The timing of delay between the two measurements is selected so they occur at the same phase of the $60\ {\rm Hz}$ mains frequency (each sequence is synchronized to the mains, so this phase is stable between repetitions).

The experimental cycle then moves on with $2.5\ {\rm s}$ of evaporative cooling, yielding $N \approx 3 \times 10^5$ atom {BEC}s at $T \lesssim 60$ nK with harmonic trap frequencies $(\omega_x,\omega_y,\omega_z) / 2 \pi \approx (31,103,155)\ {\rm Hz}$.

We diagnose the performance of our field locking using Ramsey interferometry consisting of a pair of $\pi/2$ microwave pulses (also on the $\ket{1,-2}$ to $\ket{2,-2}$ transition) spaced by a \SI{100}{\micro\second} free evolution time.
The microwave oscillator frequency is blue-detuned by $1.35\ {\rm kHz}$ to yield near-equal final populations in $F=1$ and $F=2$ at the target resonance condition.
After the second microwave pulse, we remove the trapping potential to initiate a $20\ {\rm ms}$ time-of-flight ({TOF}), during which a magnetic field gradient spatially separates $m_F$ sub-levels according to their magnetic moment via the Stern-Gerlach effect. 
The final spatial distribution is then measured by resonant absorption imaging; a transverse repump beam transfers any population in $F=1$ to $F=2$ for imaging.

Once each experimental run is complete, $N_1$ and $N_2$ are obtained from the {PTAI} data; we rapidly compute $\epsilon$ and obtain the control value $C$ as described in Sec.~\ref{sec:noise_aware}.
The hardware instructions setting the bias field are then updated for the next shot, which begins immediately.
A reference implementation of our complete magnetometry and feedback loop for the Python-based Labscript control system~\cite{Starkey2013} is available as Supplemental Material~\cite{supp}.

\subsection{Analysis and calibration}
\label{sec:image_analysis}

The number of atoms transferred by each pulse is obtained from absorption images of the 2D column density $\rho_{\rm 2D}(x,y)$ (for simplicity, we omit the spatial variables in what follows).
This density can be obtained\cite{Reinaudi2007,Hueck2017} from the optical depth
\begin{align*} 
\text{OD} &\equiv - \ln\left( \frac{I_{\rm a}-I_{\rm b}}{I_{\rm p} - I_{\rm b}} \right),
\end{align*}
which quantifies the attenuation of the probing optical field by the atoms in terms of the intensity with no atoms present $I_{\rm p}$, that with atoms present $I_{\rm a}$, and the background intensity without the probe beam $I_{\rm b}$.
The column density is computed by
\begin{align*} 
\rho_{\rm 2D} &= \frac{1}{\sigma_0} \left[{\rm OD} + \frac{I_{\rm p}}{I_{\rm sat}} e^{-{\rm OD}}\right] \, ,
\end{align*}
where $\sigma_0 = 3 \lambda^2 / (2\pi)$ is the resonant absorption cross-section; $\lambda \approx 780\ {\rm nm}$ is the probe laser wavelength; and $I_{\rm sat}$ is the saturation intensity.
We generally operate with ${\rm OD} \lesssim 1$, and select a high intensity of $I_{\rm p} \approx 3 I_{\text{sat}}$ to optimize the signal to noise ratio~\footnote{The data shown in Figs.~\ref{fig:R} and ~\ref{fig:twopulse} were taken at much lower intensity, where larger optical depths are artificially suppressed.  We empirically correct for this via $\text{OD}^\prime = \text{OD} + a(\text{OD})^3$, with the calibrated factor $a = 1.25$.}.
Furthermore, we employ principal component analysis ({PCA}) based probe reconstruction~\cite{Abdi2010,Jolliffe2016,BillingtonProbeReconstruction} to reduce noise from imaging artifacts as well as photoelectron shot noise.

We then obtain  $N_1$ and $N_2$ by integrating the corresponding column densities in the region where atoms are expected.
While not performed in this work, a second layer of {PCA} filtering could be used on the column densities to further reduce noise.

The Kalman filter-lock [Eqs.~\eqref{eq:kalman} and \eqref{eq:kalman_sigma}] requires an estimate of the uncertainty $\delta\epsilon$ of the error signal.
This uncertainty then derives from the intrinsic number uncertainties $\delta N_{1,2}$ via Eq.~\eqref{eq:sigma_epsilon}. 
These can be estimated by measuring $N_{1,2}$ using the shortest available pulse duration ($t=$\SI{12}{\micro\second} in our case) to minimize the effects of magnetic field noise.
At $\femax = 0.02$, this yielded $\delta N_i / N_0 \approx 0.01$, giving $\delta\epsilon \approx 0.014$ near $\Delta=0$.
In addition, the required calibration between current (A) and detuning (Hz) was obtained by scanning the microwave oscillator frequency $\omega_0/(2\pi)$ over a range of $30\ {\rm kHz}$, and finding the current that yields the $\Delta=0$ resonant condition.

\subsection{Benchmark: Ramsey interferometry}
\label{sec_stability}

\begin{figure}
\begin{center}
\includegraphics[scale=1]{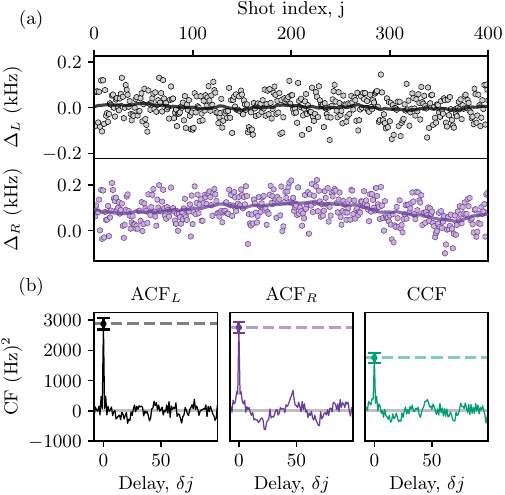}
\caption[]{
Lock performance.
(a) Detuning $\Delta$ measured with the lock active.
Top: $\Delta_L$ from {PTAI} magnetometry [also shown in Fig.~\ref{fig:intro}(c)]; bottom:  $\Delta_R$ from {TOF} Ramsey interferometry.
Markers are experimental data and the solid curve is low-pass filtered.
(b) Auto-correlation and cross-correlation functions of the data presented in (a).
}
\label{fig:autocorr}
\end{center}
\end{figure}

We now quantify the closed-loop performance of our field-lock by cross-checking against TOF Ramsey interferometry measurements.
Figure~\ref{fig:autocorr}(a) compares the detuning $\Delta_L$ measured with {PTAI} magnetometry used for locking (top, gray symbols) against the detuning $\Delta_R$ measured via Ramsey interferometry (bottom, violet symbols).
In both cases, we apply a low pass filter (solid curves) to highlight long-term drift.
As one would expect, the feedback loop's error signal $\Delta_L$ simply fluctuates about zero, however, the Ramsey signal $\Delta_R$ is offset from zero by $\approx 100\ {\rm Hz}$ and exhibits long-term drift with $\approx 20\ {\rm Hz}$ standard deviation.
While we have identified no concrete explanation for either effect, the  $2.5\ {\rm s}$ time-delay between the measurement of  $\Delta_L$ and $\Delta_R$ proves ample time for the magnetic field to drift.

Figure~\ref{fig:autocorr}(b) continues by plotting the auto-correlation functions (ACFs) of $\Delta_L$ and $\Delta_R$ (${\rm ACF}_L$ and ${\rm ACF}_R$, respectively) as well as their cross-correlation function (CCF), all computed with the low-pass filtered data subtracted.
Using the fact that the ACF evaluated at delay $\delta j=0$, i.e., ${\rm ACF}^{(0)}$, is the noise-variance, we find that both methods show a standard deviation of $\approx 53\ {\rm Hz}$; this combines the contributions of detuning noise with measurement noise.
Because the measurement noise present in $\Delta_L$ and $\Delta_R$ is statistically independent, ${\rm CCF}^{(0)}$ isolates the variance of the detuning noise, giving a standard deviation $\delta \Delta \approx 41(5)\ {\rm Hz}$, or $\delta B \approx 1.8(2)$ nT.
With the lock disabled [brown symbols in Fig.~\ref{fig:intro}(c)], the correlation analysis yields indistinguishable results at $\delta B \approx 2.0(2)$ nT.
Together with the reduced variance in ${\rm CCF}^{(0)}$, this demonstrates that the Kalman filter -- as intended -- imprints no detectable noise from the error signal onto the detuning, while still suppressing low-frequency drift.

Away from $\delta j = 0$: ${\rm ACF}_L$ is nearly structureless, with a hint of a dip between $\delta j =0$ and $50$; ${\rm ACF}_R$ exhibits clear oscillations; and the CCF again has a visible dip.
While these features could likely be mitigated by changing the selection of the loop gain $I$, they are all dwarfed by the $\delta j=0$ peak and are not relevant in practice. 

\section{Conclusion and outlook}\label{sec:conclusion}

We described a technique employing an ultracold atomic gas as an in-situ magnetometer, operating on a time scale much faster than our $\approx 15\ {\rm s}$ experimental cycle time.
Because the magnetometer measures the detuning from an atomic resonance condition, this method can detect shifts from the ac-Stark and ac-Zeeman effects, or any other physical process that detunes the targeted transition.
The magnetometer provides an error signal that we employed to stabilize the magnetic field; owing to technical limitations, our implementation updated the field once per each $\approx 15\ {\rm s}$ experimental cycle, therefore allowing us to stabilize only slow near-DC drifts of the field.

A straightforward modification of our control-system would enable feedback within the same experimental cycle.
This extension would inherently suffer from Dick sampling errors due to the cyclic nature of cold-atom experiments.
While these were not pertinent in our demonstration---because we focused on the detuning only at the sampled times---it would be present for the detuning inferred at all intervening times.

Our lock analysis focused only on the linear regime of the error signal with respect to detuning.
Modest excursions outside this regime degrade the lock's performance, while large excursions can drive the feedback loop out of lock.
More sophisticated stabilization schemes such as Bayesian parameter estimation~\cite{Carlin1997,Jaakkola2001,Schoot2021} may perform well over a wider range by directly using both measurements of number $N_1$ and $N_2$ rather than our aggregated error signal $\epsilon$.
Furthermore, we studied only simple square pulses; going beyond this, pulse shaping and composite pulse sequences can tailor the error function\cite{Levitt1986,Kenneth2004,Torosov2011}, potentially optimizing performance over a wider range of detunings or correcting for technical limitations.
For example, systematic error in the pulse area can be eliminated by replacing a single pulse with a phase-engineered sequence of multiple pulses~\cite{Torosov2019}.
In addition, the compromise between sensitivity and dynamic range can be overcome by repeating the magnetometry sequence several times in the same shot, with a range of pulse times.

Although our {PTAI}-based measurement scheme is minimally destructive, it has several side-effects beyond simply depleting atoms.
For example, any spatial structure in the probe beam briefly applies a (possibly spin-dependent) potential to the cold-atoms via the ac-Stark effect, thereby exciting the system.
Furthermore, when applied to a {BEC}, the removal of atoms also reduces the chemical potential, potentially inducing collective excitations.
While our implementation used {PTAI}, any other field-sensitive weak measurement should perform similarly; for example, Ref.~\onlinecite{Jasperse2017} used Faraday rotation, and further operated at a magic wavelength with carefully selected polarization to mitigate all ac-Stark shift effects.

\begin{acknowledgments}
The authors thank J. Argüello-Luengo and E. Anisimovas for carefully reading the manuscript.
We acknowledge Hsin-I~Lu for an early reference implementation of field locking using {PID} logic and A.~Farolfi for helpful discussions.
This work was partially supported by the National Institute of Standards and Technology; the National Science Foundation through the Quantum Leap Challenge Institute for Robust Quantum Simulation (OMA-2120757); and the Air Force Office of Scientific Research Multidisciplinary University Research Initiative ``RAPSYDY in Q'' (FA9550-22-1-0339).
E.G. gratefully acknowledges support from the Research Council of Lithuania (S-ITP-24-6).
\end{acknowledgments}

\section*{Data availability}

The data that support the findings of this article are openly available within the Supplemental Material~\cite{supp}.

%
%

\appendix

\section{Microwave transitions for rubidium-87}
\label{microwave_transitions}

\begin{figure}
\begin{center}
\includegraphics[scale=1]{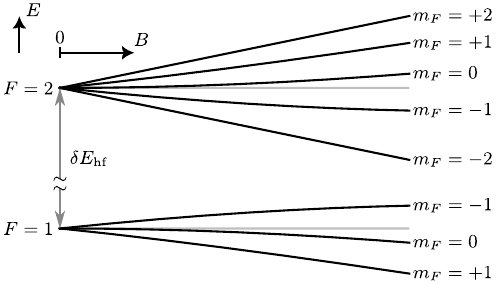}
\caption[]{Energy diagram of the $\Rb87$ ground state vs magnetic field $B$ (not to scale).
}
\label{fig:energy}
\end{center}
\end{figure}

For states with $J=1/2$ in the ground state manifold of the D transition, the Zeeman energies $E$ are found analytically from the Breit-Rabi formula~\cite{PhysRev.38.2082.2}
\begin{align}
E(B,F,m_F) =& \, \frac{-\delta E_{\text{hf}}}{2(2I+1)} + g_I \mu_B m_F B \nonumber \\
& \pm \frac{\delta E_{\text{hf}}}{2} \sqrt{ 1 + \frac{4 m_F x}{2I + 1} + x^2 } \, , \label{eq:Breit_Rabi}
\end{align}
where the $\pm$ term is $+(-)$ for $F=2(F=1)$, $\delta E_{\text{hf}} = A_{\text{hf}}(I + 1/2)$ is the zero-field hyperfine splitting between $F=2$ and $F=1$, $A_{\text{hf}}$ is the hyperfine interaction, and $x = (g_J - g_I)\mu_BB/\delta E_{\text{hf}}$.
For $\Rb87$, $\delta E_{\text{hf}}/h \approx 6.8346826109$ GHz.

The resonant oscillator frequency for a particular transition $\ket{F,m_F} \leftrightarrow \ket{F^\prime,m^\prime_F}$ is
\begin{equation}
\omega_{eg}(B) = \big| E(B,F^\prime,m^\prime_F) - E(B,F,m_F) \big| / \hbar \, .
\end{equation}
To link the detuning $\Delta$ to the magnetic field shift $\delta B$, we express the field as $B = B_0 + \delta B$,
where $B_0$ is fixed, such that the detuning is 
\begin{equation}
\Delta(B_0 + \delta B)=\omega_{eg}(B_0 + \delta B)-\omega_{eg} (B_0) \, .
\end{equation}
We then define the (local) magnetic sensitivity via fifth-order polynomial fit to $\Delta$ with $B_0$ held constant.
The linear term of this fit $\alpha_1$ is listed in Table~\ref{sensitivity_table} for two average field values: a weak field $B_0=\SI{80}{\micro\tesla}$ relevant to this work, and a much stronger field $B_0=\SI{0.2}{\tesla}$.
In the former case, the most sensitive to $B$ are the $\ket{1,\pm1} \leftrightarrow \ket{2,\pm2}$ transitions.
Increasing $B_0$ makes $\ket{1,+1} \leftrightarrow \ket{2,+2}$ more sensitive to changes in the field, while the opposite is true for $\ket{1,-1} \leftrightarrow \ket{2,-2}$.

\begin{table}
    \centering
    \caption{Resonant oscillator frequency $\omega_{eg}(B_0)=\delta E_{\text{hf}}/\hbar+\delta \omega_{eg}(B_0)$ and the linear magnetic sensitivity $\alpha_1$ of microwave transitions in $\Rb87$ for relatively weak magnetic fields.
    Boldface denotes the transition used in the present work.
    }
    \label{sensitivity_table}

    \renewcommand{\arraystretch}{1.2}
    \setlength{\tabcolsep}{6pt}
    \resizebox{\columnwidth}{!}{
        \begin{tabular}{ c  r r  r r } 
            \hline\hline
            Transition & $\delta \omega_{eg}/2\pi$, & $\alpha_1/2\pi$, & $\delta \omega_{eg}/2\pi$, & $\alpha_1/2\pi$, \\
            & (MHz) & (kHz/\SI{}{\micro\tesla}) & (MHz) & (kHz/\SI{}{\micro\tesla}) \\ \hline
            & \multicolumn{2}{c}{$B_0=$ \SI{80}{\micro\tesla}} & \multicolumn{2}{c}{$B_0=$ \SI{0.2}{\tesla}}\\\hline
            $\ket{1,+1} \leftrightarrow \ket{2,+2}$ & 1.68 & 21.02 & 4780 & 25.73 \\ 
            $\mathbf{\ket{1,-1} \leftrightarrow \ket{2,-2}}$ & $\mathbf{-1.68}$ & $\mathbf{-21.01}$ & -3063 & -9.14  \\
            $\ket{1,+1} \leftrightarrow \ket{2,+1}$ & 1.12 & 14.03 & 3958 & 23.45 \\
            $\ket{1,-1} \leftrightarrow \ket{2,-1}$ & -1.12 & -14.01 & -523 & 9.73 \\
            $\ket{1,+1} \leftrightarrow \ket{2, \hspace{2.8mm} 0}$ & 0.56 & 7.03 & 2985 & 20.63 \\
            $\ket{1,-1} \leftrightarrow \ket{2, \hspace{2.8mm} 0}$ & -0.56 & -7.02 & 739 & 13.74 \\
            $\ket{1, \hspace{2.8mm} 0} \leftrightarrow \ket{2,+1}$ & 0.56 & 7.00 & 2979 & 20.60 \\
            $\ket{1, \hspace{2.8mm} 0} \leftrightarrow \ket{2,-1}$ & -0.56 & -6.99 & 744 & 13.77 \\
            $\ket{1, \hspace{2.8mm} 0} \leftrightarrow \ket{2, \hspace{2.8mm} 0}$ & 4E-4 & 0.01 & 2006 & 17.79 \\
            \hline\hline
        \end{tabular}
    }
\end{table}

\section{Ensemble depletion}
\label{app:depletion}

\begin{table}
    \centering
    \caption{
    Examples of optimal control parameters for $t=\SI{100}{\micro\second}$ and $\delta\omega=\DeltaMax$.
    }
    \label{table_pulse}
    
    \renewcommand{\arraystretch}{1.2}
    \setlength{\tabcolsep}{6pt}
    \resizebox{\columnwidth}{!}{
        \begin{tabular}{c c c c c} 
            \hline\hline
            
            $f_0$ & $\Omega/2 \pi\ ({\rm Hz})$ & $\DeltaMax/2 \pi\ ({\rm Hz})$ & $1/(2 \pi \R)   ({\rm Hz})$ & $\deterr/2\pi\ ({\rm Hz})$ \\
            \hline
            0.001 & 96.20  & 4147.73  & 3226.63  & -0.81  \\
            0.003 & 166.72  & 4147.50  & 3224.74  & -2.42   \\
            0.01 & 304.99 & 4146.72  & 3218.12  & -8.09    \\
            0.03  & 531.29  & 4144.44  & 3199.06  & -24.36   \\
            0.1  & 990.58  & 4135.98  & 3130.65  & -82.40  \\
            0.3  & 1841.48  & 4106.37  & 2917.62  & -258.13  \\
            \hline\hline
        \end{tabular}
    }
\end{table}

Here, we characterize the two-pulse magnetometer without taking the small-transfer approximation $\femax \to 0$.
In that case, the ensemble is depleted by the first pulse, reducing the second fractional transfer.
This can significantly modify the error signal $\epsilon$, which we investigate in the following.

The exact relationship between $\femax$ and the pulse area $t \Omega$ is found by inverting Eq.~\eqref{eq:Rabi_oscillations} for $\Delta=0$:
\begin{equation}
t \Omega = 2 \arcsin{\sqrt{\femax}} \, .
\end{equation}
The total fraction $f_0=f_1+(1-f_1)f_2$ transferred by the two-pulse magnetometer for $\Delta=0$ and any $\delta\omega$ is
\begin{equation}
f_0 \approx 8 \femax \sin^2{\left( \frac{\beta}{2}  \right)} \frac{(t \delta\omega)^2 + \femax \left( 3 + \cos^2{\beta} \right)}{\beta^4}  \, ,
\end{equation}
with $\beta = \sqrt{4\femax + (t \delta\omega)^2} $.
Here, $f_1$ and $f_2$ are given by Eq.~\eqref{eq:Rabi_oscillations}, each bounded by $\femax$ as determined by $t \Omega$.

The error signal given by Eq.~\eqref{epsilon_definition} is fully described by
\begin{equation} \label{epsilon_higher_order}
\epsilon = \frac{N_1 - N_2}{N_1 + N_2} = \frac{f_1-(1-f_1)f_2}{f_1+(1-f_1)f_2}\, .
\end{equation}
Given $f_1=f_2$, $\epsilon \to 0$ and $\epsilon \to 1$ with $f_1 \to 0$ and $f_1 \to 1$, respectively; in practice, the two-pulse magnetometry scheme is used in-between these two limits.
We now describe corrections to the quantities of interest introduced in the main text by taking into account finite $\femax$.

The optimal spacing $\delta\omega$ for the "linear" regime (Fig.~\ref{fig:regimes} column 1) is found to be
\begin{multline}
t \delta \omega/2\pi \approx 0.2043 \\
- \femax(0.155-0.225\femax-1.16\femax^2) \, , 
\end{multline}
while for the continuous error function regime (Fig.~\ref{fig:regimes} column 3) we find
\begin{equation}
t \delta \omega/2\pi \approx 0.5 + \femax(0.0125 + 0.0067\femax) \, .
\end{equation}
The above expressions are highly accurate for $\femax < 0.3$.

\subsection{Number stability regime}

Now, we explore the sensitivity regime $\delta \omega =\DeltaMax$ [Fig.~\ref{fig:regimes} column 2] for $\femax<0.5$. Eq.~\eqref{eq:Delta_p_t_p} is found to be
\begin{equation} \label{delta_d_poly}
t\DeltaMax / 2 \pi \approx 0.4147838 -\femax(0.012+0.0065\femax) \, ,
\end{equation}
while the fraction transferred by a single pulse detuned by $\DeltaMax$ is
\begin{multline}
\fe(\DeltaMax) \approx \\
\femax \left(0.5477-0.0039\femax-0.0023\femax^2 \right) \, .
\end{multline}
Next, the responsivity $R$ given by Eq.~\eqref{eq_R_small} is more accurately represented by
\begin{multline} \label{PID_P_moreterms}
\R = \frac{\partial \epsilon}{\partial \Delta}\Big|_{\Delta \to 0} \approx \\
\frac{3.0983 + \femax \left( 0.99 + 0.31\femax + 0.18\femax^2 \right)}{2 \pi / t} \, ,
\end{multline}
here the correction becomes significant with finite $\femax$.
Moreover, the resonant imbalance is found to be
\begin{multline}
\epsilon_0 \equiv \epsilon(\Delta=0) \approx \\ \femax(0.2739+0.0723\femax+0.022\femax^2) \, ,
\end{multline}
On the other hand, the systematic detuning shift $\deterr \equiv \Delta(\epsilon=0)$ is
\begin{multline}
\frac{\deterr}{2\pi} \approx \\
- \frac{\femax}{t}\left(88.4-4.6\femax-7.5\femax^2\right)\!\times\!10^{-3}\, .
\end{multline}
Both quantities tend to 0 with $\femax \to 0$: for $t=\SI{100}{\micro\second}$ and $\femax=0.01$,  $\epsilon_0 \approx 0.003$ and $\deterr \approx - 2 \pi \times 9\ {\rm Hz}$.
The most direct way to account for this shift is to adjust the microwave frequencies by $\deterr$.
Alternatively, the depletion can be compensated by selecting parameters that lead to equal transfer at the lock-point.

\bibliography{bibliography}

@article{Wintersperger2023,
	author = {Wintersperger, Karen and Dommert, Florian and Ehmer, Thomas and Hoursanov, Andrey and Klepsch, Johannes and Mauerer, Wolfgang and Reuber, Georg and Strohm, Thomas and Yin, Ming and Luber, Sebastian},
	journal = {EPJ Quantum Technol.},
	number = {1},
	pages = {32},
	title = {Neutral atom quantum computing hardware: performance and end-user perspective},
	volume = {10},
    URL={https://doi.org/10.1140/epjqt/s40507-023-00190-1},
	year = {2023}}

@article{PhysRev.38.2082.2,
  title = {Measurement of Nuclear Spin},
  author = {Breit, G. and Rabi, I. I.},
  journal = {Phys. Rev.},
  volume = {38},
  issue = {11},
  pages = {2082--2083},
  numpages = {0},
  year = {1931},
  month = {Dec},
  publisher = {American Physical Society},
  doi = {10.1103/PhysRev.38.2082.2},
}

@article{ramanathanPartialtransferAbsorptionImaging2012,
  title = {Partial-Transfer Absorption Imaging: {{A}} Versatile Technique for Optimal Imaging of Ultracold Gases},
  shorttitle = {Partial-Transfer Absorption Imaging},
  author = {Ramanathan, Anand and Muniz, S{\'e}rgio R. and Wright, Kevin C. and Anderson, Russell P. and Phillips, William D. and Helmerson, Kristian and Campbell, Gretchen K.},
  year = {2012},
  month = aug,
  volume = {83},
  pages = {083119},
  issn = {0034-6748},
  doi = {10.1063/1.4747163},
  journal = {Rev. Sci. Instrum.},
  number = {8}
}

@article{serokaRepeatedMeasurementsMinimally2019a,
  title = {Repeated Measurements with Minimally Destructive Partial-Transfer Absorption Imaging},
  author = {Seroka, Erin Marshall and Curiel, Ana Vald{\'e}s and Trypogeorgos, Dimitrios and Lundblad, Nathan and Spielman, Ian B.},
  year = {2019},
  month = dec,
  volume = {27},
  pages = {36611--36624},
  issn = {1094-4087},
  doi = {10.1364/OE.27.036611},
  copyright = {\&\#169; 2019 Optical Society of America},
  journal = {Opt. Express},
  number = {25}
}

@article{mordiniSingleshotReconstructionDensity2020a,
  title = {Single-Shot Reconstruction of the Density Profile of a Dense Atomic Gas},
  author = {Mordini, C. and Trypogeorgos, D. and Wolswijk, L. and Lamporesi, G. and Ferrari, G.},
  year = {2020},
  month = sep,
  volume = {28},
  pages = {29408--29418},
  publisher = {{Optical Society of America}},
  issn = {1094-4087},
  doi = {10.1364/OE.397567},
  copyright = {\&\#169; 2020 Optical Society of America},
  journal = {Opt. Express},
  number = {20}
}

@article{Farolfi2019,
  title = {Design and Characterization of a Compact Magnetic Shield for Ultracold Atomic Gas Experiments},
  author = {Farolfi, A. and Trypogeorgos, D. and Colzi, G. and Fava, E. and Lamporesi, G. and Ferrari, G.},
  year = {2019},
  month = nov,
  volume = {90},
  pages = {115114},
  issn = {0034-6748},
  doi = {10.1063/1.5119915},
  journal = {Rev. Sci. Instrum.},
  number = {11}
}

@article{PhysRevLett.116.200402,
  title = {Geometrical Pumping with a {Bose-Einstein} condensate},
  author = {Lu, H.-I and Schemmer, M. and Aycock, L. M. and Genkina, D. and Sugawa, S. and Spielman, I. B.},
  journal = {Phys. Rev. Lett.},
  volume = {116},
  issue = {20},
  pages = {200402},
  numpages = {5},
  year = {2016},
  month = {May},
  publisher = {American Physical Society},
  doi = {10.1103/PhysRevLett.116.200402},
}

@article{Trypogeorgos2018,
  title = {Synthetic Clock Transitions via Continuous Dynamical Decoupling},
  author = {Trypogeorgos, D. and {Vald{\'e}s-Curiel}, A. and Lundblad, N. and Spielman, I. B.},
  year = {2018},
  month = jan,
  volume = {97},
  pages = {013407},
  doi = {10.1103/PhysRevA.97.013407},
  journal = {Phys. Rev. A},
  number = {1}
}

@article{LeBlanc_2013,
   title={Direct observation of zitterbewegung in a {Bose–Einstein} condensate},
   volume={15},
   ISSN={1367-2630},
   DOI={10.1088/1367-2630/15/7/073011},
   number={7},
   journal={New J. Phys.},
   publisher={IOP Publishing},
   author={LeBlanc, L J and Beeler, M C and Jiménez-García, K and Perry, A R and Sugawa, S and Williams, R A and Spielman, I B},
   year={2013},
   month=jul, pages={073011} }

@article{Lin.Rapid.PhysRevA.79.063631,
  title = {Rapid production of $^{87}\text{R}\text{b}$ {Bose-Einstein} condensates in a combined magnetic and optical potential},
  author = {Lin, Y.-J. and Perry, A. R. and Compton, R. L. and Spielman, I. B. and Porto, J. V.},
  journal = {Phys. Rev. A},
  volume = {79},
  issue = {6},
  pages = {063631},
  numpages = {8},
  year = {2009},
  month = {Jun},
  publisher = {American Physical Society},
  doi = {10.1103/PhysRevA.79.063631},
  url = {https://link.aps.org/doi/10.1103/PhysRevA.79.063631}
}

@article{Dedman2007,
    author = {Dedman, C. J. and Dall, R. G. and Byron, L. J. and Truscott, A. G.},
    title = {Active cancellation of stray magnetic fields in a {Bose-Einstein} condensation experiment},
    journal = {Rev. Sci. Instrum.},
    volume = {78},
    number = {2},
    pages = {024703},
    year = {2007},
    month = {02},
    issn = {0034-6748},
    doi = {10.1063/1.2472600},
}

@article{Merkel2019,
    author = {Merkel, B. and Thirumalai, K. and Tarlton, J. E. and Schäfer, V. M. and Ballance, C. J. and Harty, T. P. and Lucas, D. M.},
    title = {Magnetic field stabilization system for atomic physics experiments},
    journal = {Rev. Sci. Instrum.},
    volume = {90},
    number = {4},
    pages = {044702},
    year = {2019},
    month = {04},
    issn = {0034-6748},
    doi = {10.1063/1.5080093},
}

@article{Xiao2020,
    author = {Xiao, Kangda and Wang, Li and Guo, Jun and Zhu, Maohua and Zhao, Xiuchao and Sun, Xianping and Ye, Chaohui and Zhou, Xin},
    title = {Quieting an environmental magnetic field without shielding},
    journal = {Rev. Sci. Instrum.},
    volume = {91},
    number = {8},
    pages = {085107},
    year = {2020},
    month = {08},
    issn = {0034-6748},
    doi = {10.1063/5.0007464},
}

@article{Zhao2023,
	author = {Zhao, M. and Restelli, A. and Tao, J. and Liang, Q. and Spielman, I. B.},
	journal = {AIP Adv.},
	month = {06},
	number = {6},
	title = {{A 20 A bipolar current source with 140 micro-A noise over 100 kHz bandwidth}},
	volume = {13},
    URL={https://doi.org/10.1063/5.0138145},
	year = {2023}}

@article{Borkowski2023,
    author = {Borkowski, Mateusz and Reichsöllner, Lukas and Thekkeppatt, Premjith and Barbé, Vincent and van Roon, Tijs and van Druten, Klaasjan and Schreck, Florian},
    title = {Active stabilization of kilogauss magnetic fields to the ppm level for magnetoassociation on ultranarrow {Feshbach} resonances},
    journal = {Rev. Sci. Instrum.},
    volume = {94},
    number = {7},
    pages = {073202},
    year = {2023},
    month = {07},
    issn = {0034-6748},
    doi = {10.1063/5.0143825},
}

@article{Chin2010,
  title = {Feshbach resonances in ultracold gases},
  author = {Chin, Cheng and Grimm, Rudolf and Julienne, Paul and Tiesinga, Eite},
  journal = {Rev. Mod. Phys.},
  volume = {82},
  issue = {2},
  pages = {1225--1286},
  numpages = {0},
  year = {2010},
  month = {Apr},
  publisher = {American Physical Society},
  doi = {10.1103/RevModPhys.82.1225},
}

@article{Bloch2008,
  title = {Many-body physics with ultracold gases},
  author = {Bloch, Immanuel and Dalibard, Jean and Zwerger, Wilhelm},
  journal = {Rev. Mod. Phys.},
  volume = {80},
  issue = {3},
  pages = {885--964},
  numpages = {0},
  year = {2008},
  month = {Jul},
  publisher = {American Physical Society},
  doi = {10.1103/RevModPhys.80.885},
}

@article{Yang2019,
    author = {Yang, Yu-Meng and Xie, Hong-Tai and Ji, Wen-Chao and Wang, Yue-Fei and Zhang, Wei-Yong and Chen, Shuai and Jiang, Xiao},
    title = {Ultra-low noise and high bandwidth bipolar current driver for precise magnetic field control},
    journal = {Rev. Sci. Instrum.},
    volume = {90},
    number = {1},
    pages = {014701},
    year = {2019},
    month = {01},
    issn = {0034-6748},
    doi = {10.1063/1.5046484},
}

@article{ODwyer2020,
    author = {O’Dwyer, Carolyn and Ingleby, Stuart J. and Chalmers, Iain C. and Griffin, Paul F. and Riis, Erling},
    title = {A feed-forward measurement scheme for periodic noise suppression in atomic magnetometry},
    journal = {Rev. Sci. Instrum.},
    volume = {91},
    number = {4},
    pages = {045103},
    year = {2020},
    month = {04},
    issn = {0034-6748},
    doi = {10.1063/5.0002964},
}

@article{Duan2022,
    author = {Duan, Zhi-Xin and Wu, Wei-Tao and Lin, Yue-Tong and Yang, Sheng-Jun},
    title = {Simple and active magnetic-field stabilization for cold atom experiments},
    journal = {Rev. Sci. Instrum.},
    volume = {93},
    number = {12},
    pages = {123201},
    year = {2022},
    month = {12},
    issn = {0034-6748},
    doi = {10.1063/5.0119778},
}

@article{Sarkany2014,
  title = {Controlling the magnetic-field sensitivity of atomic-clock states by microwave dressing},
  author = {S\'ark\'any, L. and Weiss, P. and Hattermann, H. and Fort\'agh, J.},
  journal = {Phys. Rev. A},
  volume = {90},
  issue = {5},
  pages = {053416},
  numpages = {6},
  year = {2014},
  month = {Nov},
  publisher = {American Physical Society},
  doi = {10.1103/PhysRevA.90.053416},
}

@article{Liu2023,
    author = {Liu, Haotian and Peng, Shuai and Jiao, Bolong and Li, Jiaming and Luo, Le},
    title = {Ultra-low noise bipolar current source for ultracold atom magnetic system},
    journal = {Rev. Sci. Instrum.},
    volume = {94},
    number = {5},
    pages = {053201},
    year = {2023},
    month = {05},
    issn = {0034-6748},
    doi = {10.1063/5.0142948},
}

@article{Rogora2024,
  title = {Progress toward a zero-magnetic-field environment for ultracold-atom experiments},
  author = {Rogora, C. and Cominotti, R. and Baroni, C. and Andreoni, D. and Lamporesi, G. and Zenesini, A. and Ferrari, G.},
  journal = {Phys. Rev. A},
  volume = {110},
  issue = {1},
  pages = {013319},
  numpages = {8},
  year = {2024},
  month = {Jul},
  publisher = {American Physical Society},
  doi = {10.1103/PhysRevA.110.013319},
}

@article{Jasperse2017,
  title = {Continuous Faraday measurement of spin precession without light shifts},
  author = {Jasperse, M. and Kewming, M. J. and Fischer, S. N. and Pakkiam, P. and Anderson, R. P. and Turner, L. D.},
  journal = {Phys. Rev. A},
  volume = {96},
  issue = {6},
  pages = {063402},
  numpages = {11},
  year = {2017},
  month = {Dec},
  publisher = {American Physical Society},
  doi = {10.1103/PhysRevA.96.063402},
}

@article{Hueck2017,
author = {Klaus Hueck and Niclas Luick and Lennart Sobirey and Jonas Siegl and Thomas Lompe and Henning Moritz and Logan W. Clark and Cheng Chin},
journal = {Opt. Express},
number = {8},
pages = {8670--8679},
publisher = {Optica Publishing Group},
title = {Calibrating high intensity absorption imaging of ultracold atoms},
volume = {25},
month = {Apr},
year = {2017},
URL={https://doi.org/10.1364/OE.25.008670}
}

@article{Reinaudi2007,
author = {G. Reinaudi and T. Lahaye and Z. Wang and D. Gu\'{e}ry-Odelin},
journal = {Opt. Lett.},
number = {21},
pages = {3143--3145},
publisher = {Optica Publishing Group},
title = {Strong saturation absorption imaging of dense clouds of ultracold atoms},
volume = {32},
month = {Nov},
year = {2007},
doi = {10.1364/OL.32.003143},
}

@article{Starkey2013,
   title={A scripted control system for autonomous hardware-timed experiments},
   volume={84},
   ISSN={1089-7623},
   URL={https://doi.org/10.1063/1.4817213},
   number={8},
   journal={Rev. Sci. Instrum.},
   publisher={AIP Publishing},
   author={Starkey, P. T. and Billington, C. J. and Johnstone, S. P. and Jasperse, M. and Helmerson, K. and Turner, L. D. and Anderson, R. P.},
   year={2013},
   month=aug }

@article{Kalman1960,
	author = {Kalman, Rudolph Emil},
	journal = {J. Basic Eng.},
	number = {Series D},
	pages = {35--45},
	title = {A New Approach to Linear Filtering and Prediction Problems},
	volume = 82,
    URL={https://doi.org/10.1115/1.3662552},
	year = 1960}

@article{Carlin1997,
author = {Carlin, Bradley P. and Louis, Thomas A.},
title = {{Bayes} and empirical {Bayes} methods for data analysis},
year = {1997},
issue_date = {1997},
publisher = {Kluwer Academic Publishers},
address = {USA},
volume = {7},
number = {2},
issn = {0960-3174},
url = {https://doi.org/10.1023/A:1018577817064},
doi = {10.1023/A:1018577817064},
journal = {Stat. Comput.},
month = jun,
pages = {153–154},
numpages = {2}
}

@article{Schoot2021,
  author = {van de Schoot, Rens and Depaoli, Sarah and King, Ruth and Kramer, Bianca and Märtens, Kaspar and Tadesse, Mahlet G. and Vannucci, Marina and Gelman, Andrew and Veen, Duco and Willemsen, Joukje and Yau, Christopher},
  title = {Bayesian statistics and modelling},
  journal = {Nat. Rev. Methods Primers},
  volume = {1},
  number = {1},
  pages = {26},
  year = {2021},
  publisher = {Nature Publishing Group},
  doi = {10.1038/s43586-020-00001-2}
}

@article{Jaakkola2001,
author = {Jaakkola, Tommi and Jordan, Michael},
year = {2000},
pages = {25--27},
title = {Bayesian Parameter Estimation Via Variational Methods},
volume = {10},
journal = {Stat. Comput.},
doi = {10.1023/A:1008932416310}
}

@article{Torosov2011,
  title = {Smooth composite pulses for high-fidelity quantum information processing},
  author = {Torosov, Boyan T. and Vitanov, Nikolay V.},
  journal = {Phys. Rev. A},
  volume = {83},
  issue = {5},
  pages = {053420},
  numpages = {7},
  year = {2011},
  month = {May},
  publisher = {American Physical Society},
  doi = {10.1103/PhysRevA.83.053420},
  url = {https://link.aps.org/doi/10.1103/PhysRevA.83.053420}
}

@article{Torosov2019,
   title={Composite pulses with errant phases},
   volume={100},
   ISSN={2469-9934},
   url={http://dx.doi.org/10.1103/PhysRevA.100.023410},
   number={2},
   journal={Phys. Rev. A},
   publisher={American Physical Society (APS)},
   author={Torosov, Boyan T. and Vitanov, Nikolay V.},
   year={2019},
   month=aug }

@article{Levitt1986,
title = {Composite pulses},
journal = {Prog. Nucl. Magn. Reson. Spectrosc.},
volume = {18},
number = {2},
pages = {61-122},
year = {1986},
issn = {0079-6565},
doi = {https://doi.org/10.1016/0079-6565(86)80005-X},
url = {https://www.sciencedirect.com/science/article/pii/007965658680005X},
author = {Malcolm H. Levitt}
}

@article{Kenneth2004,
  title = {Arbitrarily accurate composite pulse sequences},
  author = {Brown, Kenneth R. and Harrow, Aram W. and Chuang, Isaac L.},
  journal = {Phys. Rev. A},
  volume = {70},
  issue = {5},
  pages = {052318},
  numpages = {4},
  year = {2004},
  month = {Nov},
  publisher = {American Physical Society},
  doi = {10.1103/PhysRevA.70.052318},
  url = {https://link.aps.org/doi/10.1103/PhysRevA.70.052318}
}

@article{Jolliffe2016,
author = {Jolliffe, Ian T.  and Cadima, Jorge },
title = {Principal component analysis: a review and recent developments},
journal = {Philos. Trans. R. Soc. A},
volume = {374},
number = {2065},
pages = {20150202},
year = {2016},
doi = {10.1098/rsta.2015.0202},
URL = {https://doi.org/10.1098/rsta.2015.0202}
}

@article{Abdi2010,
author = {Abdi, Hervé and Williams, Lynne J.},
title = {Principal component analysis},
journal = {WIREs Comput. Stat.},
volume = {2},
number = {4},
pages = {433-459},
doi = {10.1002/wics.101},
url = {https://doi.org/10.1002/wics.10},
year = {2010}
}

@article{Ana2017,
doi = {10.1088/1367-2630/aa6279},
url = {https://doi.org/10.1088/1367-2630/aa6279},
year = {2017},
month = {mar},
publisher = {IOP Publishing},
volume = {19},
number = {3},
pages = {033025},
author = {Valdés-Curiel, A and Trypogeorgos, D and Marshall, E E and Spielman, I B},
title = {Fourier transform spectroscopy of a spin–orbit coupled {Bose} gas},
journal = {New J. Phys.},
}

@article{Lin2011,
   title={Spin–orbit-coupled {Bose–Einstein} condensates},
   volume={471},
   ISSN={1476-4687},
   url={http://dx.doi.org/10.1038/nature09887},
   DOI={10.1038/nature09887},
   number={7336},
   journal={Nature},
   publisher={Springer Science and Business Media LLC},
   author={Lin, Y.-J. and Jiménez-García, K. and Spielman, I. B.},
   year={2011},
   month=mar, pages={83–86} 
}

@article{Ludlow2015,
  title = {Optical atomic clocks},
  author = {Ludlow, Andrew D. and Boyd, Martin M. and Ye, Jun and Peik, E. and Schmidt, P. O.},
  journal = {Rev. Mod. Phys.},
  volume = {87},
  issue = {2},
  pages = {637--701},
  numpages = {65},
  year = {2015},
  month = {Jun},
  publisher = {American Physical Society},
  doi = {10.1103/RevModPhys.87.637},
  url = {https://link.aps.org/doi/10.1103/RevModPhys.87.637}
}

@article{Saffman2016,
  author = {M. Saffman},
  title = {Quantum computing with atomic qubits and {Rydberg} interactions: progress and challenges},
  journal = {J. Phys. B: At. Mol. Opt. Phys.},
  publisher = {{IOP} Publishing},
  year = {2016},
  volume = {49},
  number = {20},
  pages = {202001},
  doi = {10.1088/0953-4075/49/20/202001},
  url = {https://doi.org/10.1088%2F0953-4075%2F49%2F20%2F202001}
}

@article{Mehlstaubler2018,
	author = {Mehlst{\"a}ubler, Tanja E and Grosche, Gesine and Lisdat, Christian and Schmidt, Piet O and Denker, Heiner},
	journal = {Rep. Prog. Phys.},
	month = {apr},
	number = {6},
	pages = {064401},
	title = {Atomic clocks for geodesy},
	volume = {81},
    URL={https://doi.org/10.1088/1361-6633/aab409},
	year = {2018}}

@misc{BillingtonProbeReconstruction,
  author = {J. Billington, Chris},
  title = {Python library for reconstructing images as linear sums of reference images},
  year = {2020},
  publisher = {GitHub},
  journal = {GitHub repository},
  howpublished = {\url{https://github.com/chrisjbillington/image_reconstruction}},
}

@misc{supp,
  note = "See Ancillary files for the data
    of the experiments and the reference implementation of the feedback loop."
}

@inproceedings{Dick1988,
	author = {G. John Dick},
	booktitle = {Proceedings of the 19th Precise Time and Time Interval (PTTI) Applications and Planning Meeting},
	pages = {133-147},
	title = {Local oscillator induced instabilities in trapped ion frequency standards},
	year = {1987}}

\end{document}